\documentclass{article}
\usepackage{setspace}
\usepackage{color}
\usepackage{graphicx}
\usepackage{fullpage, amsmath, amssymb}

\pdfoutput=0

\begin{document}

\begin{center}
\subsection*{Finding bridges in packings of colloidal spheres}

Matthew C. Jenkins$^{1,2,\star}$, Mark D. Haw$^{1,3}$, Gary C. Barker$^{4}$, Wilson C. K. Poon$^{1}$, Stefan U. Egelhaaf$^{1,2}$\\
\medskip
$^1$School of Physics and Astronomy and COSMIC, The University of Edinburgh, James Clerk Maxwell Building, Kings Buildings, Mayfield Road, Edinburgh, EH9 3JZ, Scotland, UK.\\
$^2$Condensed Matter Physics Laboratory, Heinrich-Heine-University, Universit\"atsstra{\ss}e 1, 40225, D\"usseldorf, Germany.\\
$^3$Department of Chemical and Process Engineering, University of Strathclyde, James Weir Building, 75 Montrose Street, Glasgow, G1 1XJ, Scotland, UK.\\
$^4$Institute of Food Research, Norwich Research Park, Colney, Norwich, NR4 7UA, UK.\\
\medskip
$^\star$corresponding author: matthew.jenkins@uni-duesseldorf.de
\end{center}

\bigskip
\bigskip
\bigskip

\section*{Abstract}
We identify putative load-bearing structures (bridges) in
experimental colloidal systems studied by confocal microscopy.
Bridges are co-operative structures that have been used to explain
stability and inhomogeneous force transmission in simulated granular
packings with a range of densities. We show that bridges similar to
those found in granular simulations are present in real experimental
colloidal packings.  We describe critically the bridge-finding
procedure for real experimental data and propose a new
criterion-Lowest Mean Squared Separation (LSQS)-for selecting
optimum stabilisations.

\newpage


\section{Introduction \label{introduction}}

Sphere packings have long been of interest: Kepler conjectured in
the seventeenth century that the maximum attainable volume fraction
for freely-placed hard spheres is $\phi = \pi/\sqrt{18}$, achieved
for close-packed crystalline arrangements \cite{Kepler1611}.  This
was confirmed only recently \cite{Hales05}.  Random sphere packings
are less well understood, despite their long-standing scientific
interest \cite{Bernal59a, Bernal64b}, wide-spread industrial
engineering relevance \cite{AsteBook}, and numerous important
realisations ranging from cannonballs to powders and sand grains in
dunes.  For a stable random packing, the particles must be confined,
possibly by caging \cite{Philipse03, Peters01}.  Maxwell argued that
for mechanical equilibrium an individual particle requires $d+1$
neighbours in $d$ dimensions, but that stability of an assembly of
particles requires on average at least $2d$ neighbours
\cite{Alexander98, Wyart05, Maxwell1864}.  (Note that for {\em
frictional} particles, only $d+1$ neighbours are needed for {\em
global} stability \cite{Edwards98, MehtaBook}.)  These requirements
can be met for volume fractions $\phi$ ranging from the so-called
random loose packed ($\phi_{\mathrm{rlp}} \approx 0.55$) to the
random close packed ($\phi_{\mathrm{rcp}} \approx 0.64$) limit
\cite{AsteBook, Onoda90, Jerkins08, Torquato00}.

The observed stability of packings in a volume fraction range from
$\phi_{\mathrm{rlp}}$ to $\phi_{\mathrm{rcp}}$ suggests that
load-bearing co-operative structures involving multiple particles
exist and provide stability against gravity \cite{Onoda90, Nolan92}.
Simulations in 2D \cite{Arevalo06} and 3D \cite{MehtaBook,
Pugnaloni04, Pugnaloni01, Mehta04, Mehta09} have identified
load-bearing structures --- bridges --- in an attempt to explain the
existence of stable packings at different volume fractions.  Bridges
can have very different sizes and extensions about an axis
\cite{Pugnaloni04} and various architectures, ranging from linear
(`string-like') bridges to more complex, branched bridges
\cite{MehtaBook, Pugnaloni01, Mehta04, Mehta09}.  In this study, we
are particularly interested in the bridge size distribution, that
is, the probability that a particle belongs to a bridge of size $m$,
$P(m) = mN(m)/N_{\mathrm{tot}}$ with $N(m)$ the number of bridges in
the packing comprising $m$ members and $N_{\mathrm{tot}}$ the total
number of bridges in the packing.

Random packings, formed due to the effect of gravity, are typically
non-ergodic and geometric frustration prevents them from attaining
the thermodynamically preferred crystalline state.  Whether there is
a connection between frustrated states in granular and colloidal
systems is not clear. While gravity dominates in granular systems,
colloids are governed by thermal energy, so that particles with a
different density from the solvent nevertheless remain dispersed due
to thermal motion.  The random thermal (Brownian) motion enables the
particles to explore different spatial configurations, leading to
ergodic states.  In dense suspensions, however, their dynamics are
strongly suppressed and metastable supercooled liquids and
non-ergodic glasses can be observed \cite{Torquato00b, Speedy94,
Speedy98a, Rintoul96, vanMegen91, vanMegen93, vanMegen93b,
vanMegen94}. The relative importance of gravity and thermal energy,
and thus whether a system is granular or colloidal, necessarily
depends only on the ratio of the gravitational and thermal energies,
characterised by the gravitational Peclet number
$\mathrm{Pe}_{\mathrm{g}} \sim m_\mathrm{B} g R/k_\mathrm{B}T$,
where $R$ is the particles' radius, $m_{\mathrm{B}}$ their buoyant
mass, $g$ the acceleration due to gravity and $k_\mathrm{B}T$ the
thermal energy.

We investigate experimentally a colloidal system of hard spheres. The samples are studied by confocal microscopy. Although this allows us to determine the particle locations with high accuracy \cite{Jenkins08}, the still finite accuracy of the determined co-ordinates and the polydispersity in particle size can lead to complications. The particle co-ordinates are used to search for bridges following concepts and methods developed in the study of granular systems \cite{MehtaBook, Pugnaloni04, Pugnaloni01, Mehta04, Mehta09}. This permits a direct comparison of the features of bridges in colloidal and granular systems.


\section{Materials and Methods \label{exptsection}}

\subsection{Samples}

We used poly-methylmethacrylate (PMMA) particles which are
sterically stabilised by poly-12-hydroxystearic acid (PHSA) and
fluorescently labelled using
4-methylaminoethylmethacrylate-7-nitrobenzo-2-oxa-1,3-diazol (NBD)
suspended in {\em cis}-decahydronaphthalene ({\em cis}-decalin).
These particles behave as nearly hard spheres \cite{PuseyLesHouches,
Pusey86}.  Static light scattering from a dilute sample indicates a
mean particle diameter of $2R=2.15\pm0.02\mu$m. This size is
consistent with the position of the first peak in the pair
correlation function $g(r)$, which was determined from the particle
co-ordinates.  Based on the mean particle radius and the average
Vorono\"{i} volume per particle, the volume fractions $\phi$ were
determined \cite{mythesis}.  The density difference between the
solvent and particles results in $Pe_\mathrm{g} \approx 4$, and
sedimentation occurs relatively rapidly; this allows us to
investigate samples of any volume fraction up to $\phi \lesssim
\phi_{\mathrm{rcp}}$.  Here we focus on several dense samples of
mean volume fraction $\phi =0.608 \pm 0.004$ to investigate features
of the bridge-finding process.

\subsection{Confocal Microscopy}
\label{confocalMicroscopy}

The samples were investigated using a fast-scanning confocal scanhead (VT-Eye, Visitech International Ltd.) connected to a Nikon Eclipse TE300 inverted microscope. This permits large regions to be captured sufficiently quickly (typically in around $3$~s) even in dilute suspensions, where the constituent particles move relatively quickly.  All of the sample volumes imaged here were $512\times512\times100\;\mu$m voxels; the lateral pixel pitch was typically around $0.13\;\mu$m~pixel$^{-1}$ (measured with a high-resolution calibration slide \cite{Jenkins08}) and the axial one was around $0.2\;\mu$m~pixel$^{-1}$.  This gives a typical imaged volume of about $70\;\mu$m~$\times\;70\;\mu$m~$\times\;20\;\mu$m.


\section{Finding Bridges}

We follow a method of bridge finding developed for the analysis of simulation data of granular systems \cite{MehtaBook, Pugnaloni04, Pugnaloni01, Mehta04}. The main feature of a bridge is that the constituent particles are both stable, that is, are prevented from falling or settling under the influence of gravity, and mutually stabilising, that is, must act co-operatively. In the following we discuss these two properties and their implementation in a bridge finding algorithm.

\subsection{Stability Criterion \label{StabilityCriterion}}

\begin{figure}[h]
\begin{center}
\includegraphics[width=15cm, clip=true, trim=0 12mm 0 0]{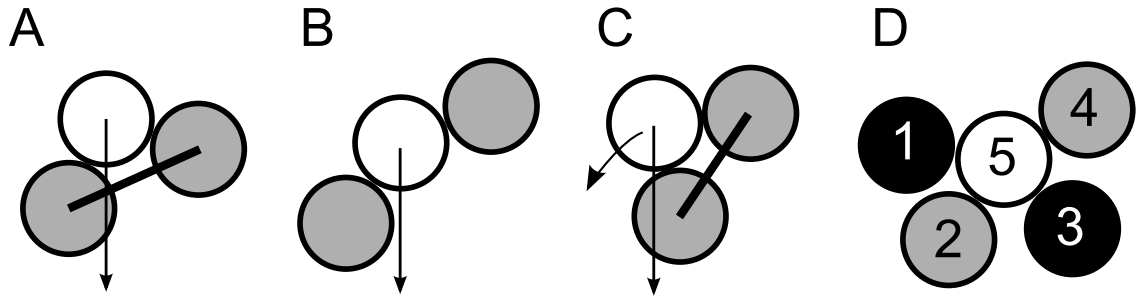}
\caption[In two dimensions, a stable particle and an unstable
particle.]{In two dimensions, (A) a stable particle (open circle) with its centre above both supporting particles (solid circles) and (B) below one of the supporting particles with, in both cases, the weight vectors between the centres of the two supporting particles. (C) A particle (open circle) with its weight vector outside the line connecting the centres of the supporting particles (solid circles) is unstable. (D) A stable particle (open circle) can be stabilised by more than one subset; the three pairs 1-3, 2-3 and 2-4 independently give support to the stable particle 5.\label{2dstabilitydiag}}
\end{center}
\end{figure}

A particle can only belong to a bridge if it is stable with respect to the applied force. 
For clarity, we consider stability in two dimensions (Fig.~\ref{2dstabilitydiag}). Each stable particle must be supported by at least two contacting neighbours, which must be arranged so that the weight vector of the candidate stable particle passes between the centres of the two stabilising or base particles (Fig.~\ref{2dstabilitydiag}A). (In three dimensions, a stable particle requires at least three contacting neighbours, with the weight vector passing through the triangle defined by the centres of the stabilising particles.) Their centres can be below the supported particle (Fig.~\ref{2dstabilitydiag}A), or one (two in three dimensions) can also be above the supported particle (Fig.~\ref{2dstabilitydiag}B). In any case, for a particle to be stable, its weight vector must not pass outside the line (triangle in three dimensions) connecting the centres (Fig.~\ref{2dstabilitydiag}C).

Importantly, we note here that there can be more than one stabilising subset per stable particle in both two and three dimensions. Fig.~\ref{2dstabilitydiag}D shows a stable particle with four stabilising particles arranged into three stabilising subsets: 1-3, 2-3 and 2-4.  This leads to redundancy in the stabilising network, which we discuss below.

In general, not all particles in a packing need be stable; even in high density samples so-called rattlers can be found. Rattlers are particles which are free to move, driven by thermal motion, within a, typically small, volume and are thus not stable against gravity.  In random packings they exist in small, protocol-dependent proportions even in the most dense randomly-packed states \cite{Torquato00}. In addition, particles can be (wrongly) declared rattlers due to polydispersity and uncertainty in the co-ordinates.

\subsection{Mutual Stabilisations\label{bridgebasicsmutstabs}}

Sphere packings are said to be stable if each member of the packing is stable against a uniaxially-applied force, such as gravity.  Sphere packings are known to be stable against gravity for a range of volume fractions; $0.55 \lesssim \phi \lesssim 0.64$ (e.g. \cite{AsteBook, Onoda90, Jerkins08, Torquato00}). For this to be possible, stability (Sec. \ref{StabilityCriterion}) is not sufficient; mutual stabilisations are required.  One particular way of grouping mutual stabilisations is by assigning them to bridges.

\begin{figure}[h]
\begin{center}
 \includegraphics[width=4.5cm,angle=0]{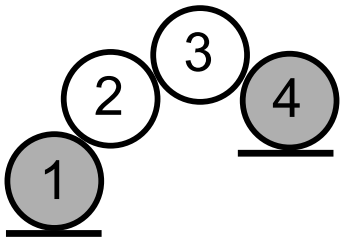}
\caption{Base (1 and 4) and mutually stabilising (2 and 3) particles in two dimensions.\label{mutstabspheres}}
\end{center}
\end{figure}

A stable particle which does not participate in any mutual stabilisations represents the simplest bridge, of size $m = 1$.  Though this is a trivial case, we still regard it as a bridge (though some results specifically exclude bridges of size $m = 1$). Longer bridges require co-operative effects, referred to as mutual stabilisations. We illustrate mutual stabilisation for two dimensions (Fig.~\ref{mutstabspheres}). The particles labelled 1 and 4 are base particles, which are necessary but rely on no members of the bridge for their own stability. The other particles (2 and 3) rely on other particles for their stability; 2 on 1 and 3, 3 on 2 and 4. For this reason, they are both stable and therefore each belongs to a bridge.  Moreover, since 2 and 3 rely on each other for stability, that is, removing particle 2 would cause particle 3 to fall and vice versa, they (particles 2 and 3) are mutually stabilising.  Base particles (here 1 and 4), although crucial in the bridge, are, by convention, not considered members of the bridge.  This is in contrast to a `real' bridge, where we would always consider the base particles to belong to the bridge; intuitively, the base particles are viewed as fixed buttresses rather than `stones' in the bridge.  This choice ensures that each stable particle belongs to one and only one bridge.  In this case, mutually stabilising particles 2 and 3 belong to a two-particle bridge.

\subsection{Algorithm\label{algorithm}}

We now describe and discuss the algorithm used for identifying
stable particles and mutual stabilisations.

\subsubsection*{Step 1 -- Find particle co-ordinates}

Using confocal microscopy (Sec.~\ref{confocalMicroscopy}), we obtain
three-dimensional micrographs of each sample. Since we are
investigating concentrated suspensions, particle separations are
small and particle images are thus frequently overlapping (see also
\cite{Baumgartl05, Baumgartl06}). This, together with noise,
necessitates extra care when determining particle co-ordinates from
the micrographs. In addressing these issues, we use an iterative
algorithm which is based on a frequently-used recipe
\cite{Crocker96}, but has been improved to increase the reliability
of the particle co-ordinates \cite{Jenkins08}.

\subsubsection*{Step 2 -- Establish which particles can be used}

Not all of the particles in the volume accessible to the confocal
microscope can be used during bridge finding, due to two distinct
`edge effects'.  The first is that co-ordinates of particles too
near to the edge of the observation volume are unreliable as a
result of potentially missing raw data \cite{Jenkins08}. Thus
particles whose centres are closer to the edge than one radius $R$
are excluded.  The second effect arises since a particle can only be
declared stable if it has at least three suitably placed neighbours
(Sec.~\ref{StabilityCriterion}), which themselves must also be
identified accurately. In the most extreme case, a particle can have
a neighbour up to one diameter, i.e.~$2R$, nearer to the edge of the
observed volume than its own position. We hence require that genuine
bulk particles have a separation from the edge of the observation
volume of $3R$ to avoid both effects. Whether during the
determination of a particle's location and stability border widths
of $R$ and $2R$, respectively, or vice versa (as in our case) are
used results in only small systematic differences.  The need for a
minimum separation from the edge of experimental datasets is in
contrast to the situation in simulations with periodic boundary
conditions, where no particles need to be excluded.

\subsubsection*{Step 3 -- Find contacting neighbours for each particle}

Stability requires sufficient and appropriately placed contacting neighbours, which can be identified by, for example, the Vorono\"{i} (or Wigner-Seitz) construction \cite{PreparataBook}.  Determining contacting particles is performed based simply on their separation; for arbitrarily well-located monodisperse spheres their centres must be separated by their diameter $2R$. The situation becomes more difficult for polydisperse particles (since the radii are not known on an individual particle basis), and where there is experimental uncertainty in the co-ordinates. This can lead to experimentally determined separations between contacting particles which are larger or smaller than their (mean) diameter. To be certain of finding all possible neighbours, we must allow for separations larger than the particle diameter, i.e.~$2cR$ with the cutoff value $c\ge 1.0$. Typically in colloidal model systems, the polydispersity is between 5\% and 10\% and the uncertainty in the co-ordinates similar, suggesting $1.07 \le c \le 1.14$. Below we will discuss the dependence of various parameters on $c$.



\begin{figure}[h]
\begin{center}
\includegraphics[width=10cm,angle=0,clip=true,trim=0 0 0 0]{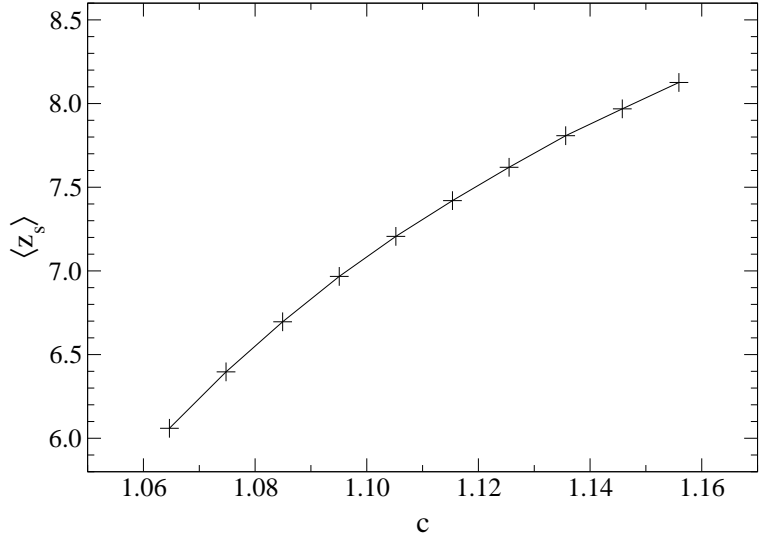}
\caption{Mean number of neighbours $\langle z_{\mathrm{s}} \rangle$ within a range defined by the cutoff value $c$. The data are based on five imaged volumes with a volume fraction $\phi=0.608 \pm 0.004$. \label{coordnowithcutoff}}
\end{center}
\end{figure}

Fig.~\ref{coordnowithcutoff} shows that the mean number of contacting neighbours $\langle z_{\mathrm{s}} \rangle$ for the potentially stable particles increases with the cutoff value $c$ as expected.

\subsubsection*{Step 4  -- Establish stabilising subsets for each particle}

If a particle is potentially stable (Step 2), every possible set of three contacting neighbours needs to be checked for its ability to provide stability. For $z$ contacting neighbours, all $^zC_3$ subsets of three particles could, in principle, be stabilising and must therefore be tested. To test a potentially stabilising subset of three particles, the weight vector of the candidate stable particle is checked for intersection with the triangle formed by the centres of the potentially stabilising particles (Fig.~\ref{2dstabilitydiag}). If this is the case, this particle and its stabilising particles are added to a list of stable particles and their stabilising subsets.


\begin{figure}[h]
\begin{center}
\includegraphics[width=10cm,angle=0,clip=true,trim=0 0 0 0]{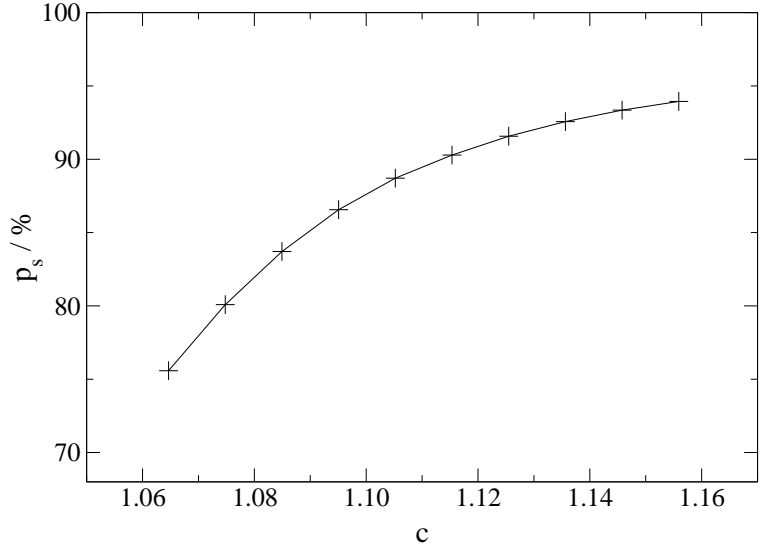}
\caption{Fraction of particles deemed stable, $p_{\mathrm{s}}$, as a function of cutoff value $c$. Sample as in figure \ref{coordnowithcutoff}. \label{stablewithcutoff}}
\end{center}
\end{figure}

The fraction of particles deemed stable in a packing, $p_{\mathrm{s}}$, depends on the cutoff value $c$ (Fig.~\ref{stablewithcutoff}). With increasing $c$, that is, with an increasing number of neighbours deemed contacting, $p_{\mathrm{s}}$ increases; this is consistent with the increase of $\langle z \rangle$ (Fig.~\ref{coordnowithcutoff}). Interestingly, the fraction of stable particles seems to saturate at $p_{\mathrm{s}} \approx 0.95$ and thus not all particles are stabilised.  This is in agreement with the small proportion of rattlers (a few percent) found in simulations of random packings \cite{OHern03}. We assume that at least some rattlers arise due to inaccuracies in the particle locations.


\begin{figure}[h]
\begin{center}
\includegraphics[width=10cm,angle=0,clip=true,trim=0 0 0 0]{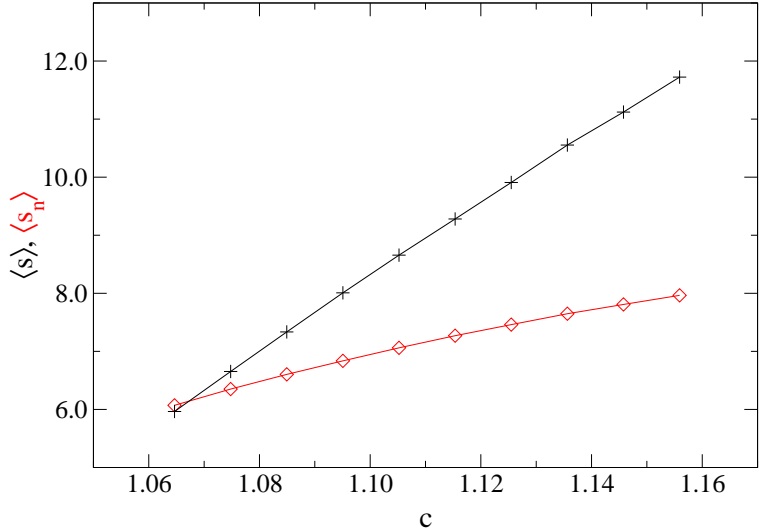}
\caption{Mean number of stabilising subsets per stable particle, $\langle s \rangle$ (black crosses), and the mean number of stabilising particles per stable particle, $\langle s_{\mathrm{n}} \rangle$ (red diamonds), both as a function of cutoff value $c$. Sample as in figure \ref{coordnowithcutoff}. \label{meanstabsubswithcutoff}}
\end{center}
\end{figure}

There is significant redundancy in the load-bearing network, Fig.
\ref{meanstabsubswithcutoff}. Many particles are stabilised by more
than one stabilising subset. The mean number of stabilising subsets
per stable particle $\langle s \rangle$ increases with increasing
$c$, as expected from the increase of $\langle z_{\mathrm{s}}
\rangle$ and $p_s$ (Figs.~\ref{coordnowithcutoff},
\ref{stablewithcutoff}).  Indeed, $\langle s \rangle$ is remarkably
high, for example around ten for $c=1.12$
(Fig.~\ref{meanstabsubswithcutoff}). This number can be compared
with the corresponding number of potentially stabilising subsets,
given by $^{\langle z_{\mathrm{s}} \rangle}\mathrm{C}_3$, with
(here) $\langle z_{\mathrm{s}} \rangle \approx 7.5$
(Fig.~\ref{coordnowithcutoff}) and thus, in this case, between a
fifth and a quarter of the subsets are stabilising subsets.  Also
shown in Fig.~\ref{meanstabsubswithcutoff} is the mean number of
stabilising {\em particles} per stable particle, $\langle
s_{\mathrm{n}} \rangle$, which is also large.  This suggests that
dense ($\phi\approx\phi_{\mathrm{rcp}}$) random sphere packings are
greatly overstabilised against uniaxial forces.

\begin{figure}[h]
\begin{center}
\includegraphics[width=10cm,angle=0,clip=true,trim=0 0 0 0]{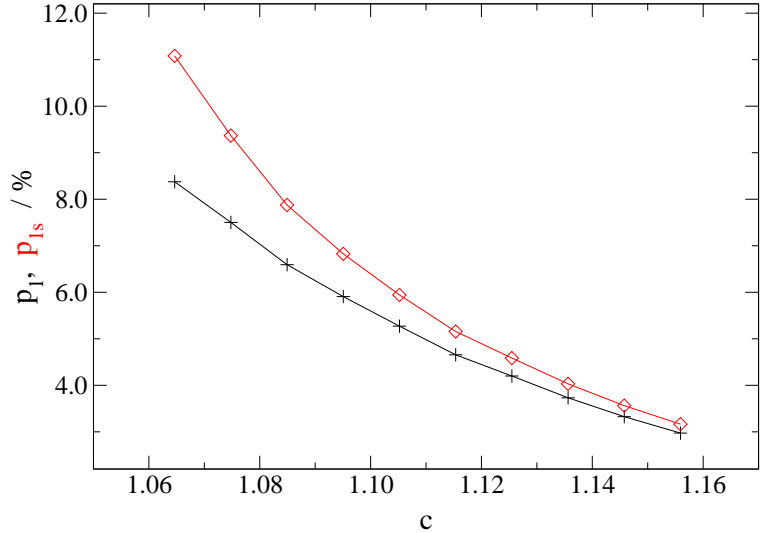}
\caption{Fraction of potentially stable particles, $p_1$ (black crosses), and actually stable particles, $p_{\mathrm{1s}}$ (red diamonds), stabilised by precisely one stabilising subset, as a function of cutoff value $c$.  Sample as in figure \ref{coordnowithcutoff}. \label{onesubsetwithcutoff}}
\end{center}
\end{figure}

Even in the most dense packings and at the highest $\langle z \rangle$ there are nevertheless stable particles which are stabilised by only one subset and are thus minimally stabilised. However, in dense samples the fraction of particles stabilised in one way, $p_1$, and (similar) the fraction of {\em stable} particles stabilised in one way, $p_{\mathrm{1s}}$, is very low, only a few percent (Fig.~\ref{onesubsetwithcutoff}).  The notion of minimal stabilisation is reminiscent of the so-called marginal rigidity state, where each stable particle is stabilised in only one way \cite{Blumenfeld05}.


\subsubsection*{Step 5 -- Choose a single stabilising subset for each particle}

To proceed we must choose, from a large number of stabilising
subsets (Fig.~\ref{meanstabsubswithcutoff}), a single subset for
each particle \cite{Arevalo06}; this is not explicitly stated in
previous publications on three-dimensional systems \cite{MehtaBook,
Pugnaloni04, Pugnaloni01, Mehta04}. Although this is a crucial
choice, there is no strict guide for its selection. (There is a
comparison of two possible criteria, lowest subset versus a
randomly-chosen one, for discs in two dimensions \cite{Arevalo06},
but none of which we are aware in three.)  The inherent experimental
uncertainty and the polydispersity of the particles, whose
individual size cannot be determined, make this decision still less
clear. We consider two options.

The first has been favoured in 3D simulations. Since simulated
packings are generated by allowing particles to fall under gravity
until stabilised, the single stabilising subset which initially
arrested the fall is known. In identifying bridges in general,
however, this information is not available.  Instead, the
stabilising subset with the lowest centre of mass (LCOM) is chosen.
It is intuitively appealing that a stabilisation by a subset with a
lower centre of mass (as shown in Fig.~\ref{2dstabilitydiag}A for
two dimensions) is `more stabilising' than one with a higher centre
of mass (Fig.~\ref{2dstabilitydiag}B). However, even if LCOM
stabilisations do coincide with the initial `real' stabilisations,
this does not preclude the creation of additional stabilisations
involving subsequently deposited spheres.  Furthermore, whether this
correspondence holds for systems which undergo significant thermal
motion is not clear.

Our colloidal systems not only undergo thermal motion, but are also polydisperse and have an uncertainty in the particle locations. As argued in Step 3, this results in an overcounting of neighbours to include some non-contacting particles, which means there are typically subsets included that cannot genuinely provide stability.  It therefore seems appropriate to choose the stabilising subset whose members are most likely to be genuinely contacting. We suggest that this is the subset whose members are on average closest to the stable particle, a choice referred to as the lowest mean squared separation  (LSQS).  We discuss the effect of the choice of the stabilising subset below (Sec.~\ref{sec:discStab}).

\subsubsection*{Step 6 -- Identify and group mutual stabilisations}

Once the stabilising subsets have been chosen, we determine all pairs of mutually stabilising particles.  A mutual stabilisation --- the defining feature of a bridge --- occurs when a particle is stabilised by a second particle which itself is stabilised by the first particle (Sec.~\ref{bridgebasicsmutstabs}).  Individual particles can participate in more than one mutually stabilising pair. We group all of the mutually stabilising pairs into disjoint clusters (using a standard algorithm \cite{Stoddard78}) so that all mutually stabilised particles belong to one but only one cluster, which represents a bridge.

The algorithm provides not only the bridges present in a packing, but also the contacting neighbours, potentially and actually stable particles, all stabilising subsets and the selected stabilising subsets as well as the mutually stabilising particle pairs. Based on this information, different packing stability and bridging properties can be derived.


\section{Interpretation}

During the bridge finding process, two steps require some choice.
One concerns the criterion for the selection of a unique stabilising
subset for each stable particle (Step 5). The other is the value of
the cutoff value $c$ used during the determination of the particles'
neighbours (Step 3). In the following we discuss the consequences of
these choices, in particular on the bridge size distribution $P(m)$.

\subsection{Choice of the stabilising subset}
\label{sec:discStab}

Each stable particle has typically several potentially stabilising
subsets; at high densities this can be as many as ten or more,
Fig.~\ref{meanstabsubswithcutoff}). From this set of potentially
stabilising subsets one has to choose a single, unique stabilising
subset during the bridge finding process. Although stability is
linked to load-bearing, the choice has to be made based on
geometrical information alone. In experimental data this geometrical
information, namely the particle location and size, is furthermore
affected by experimental uncertainties in the particle co-ordinates
and polydispersity in size. We have presented two criteria to choose
the stabilising subset (Step 5). First, the subset with the lowest
centre of mass (LCOM) is typically chosen in simulations, where
experimental uncertainties and polydispersity are absent. Second, we
propose selecting the stabilising subset whose members are on
average closest to the stable particle, the lowest mean squared
separation  (LSQS) criterion. This criterion favours the subset
whose members are most likely to be genuinely contacting, and thus
compensates for the intrinsic experimental uncertainties. The
problem of which subset to choose is far from clearly solved, and
has not been widely discussed in the literature \cite{MehtaBook,
Pugnaloni04, Pugnaloni01, Mehta04, Mehta09}.  In two dimensions, one
simulation study in which the stabilising pair that arrested a given
stable particle's fall is known found that randomly-chosen
stabilising pairs and lowest stabilising pairs provide good results
for disordered packings, with the randomly-chosen ones more
faithfully reproducing the genuine bridge size distribution
\cite{Arevalo06}.

\begin{figure}[h]
\begin{center}
\includegraphics[width=10cm,clip=true,trim=0 0 0 0]{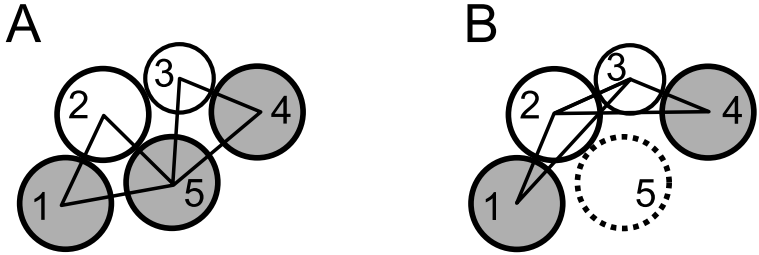}
\caption{Arrangement of polydisperse particles with the stabilising subset chosen according to the lowest centre of mass (LCOM) criterion (A) and the lowest mean squared separation  (LSQS) criterion (B). Applying the LCOM criterion (A), there are two one-particle bridges (2 stabilised by 1 and 5; 3 stabilised by 4 and 5) and three (independently stabilised) base particles (1, 4, and 5).  Based on the LSQS criterion (B), particle 2 is stabilised by 1 and 3, while particle 3 is stabilised by 2 and 4. Particles 2 and 3 are thus mutually stabilising and form a two-particle bridge. Particles 1 and 4 are base particles and particle 5, despite contacting particle 2, does not participate at all in this bridge. In this case, the LSQS criterion successfully deals with the non-contacting neighbours 3 and 5, unavoidably deemed neighbours by the necessarily ``too large'' value of $c$ (see text).\label{choiceofsubset}}
\end{center}
\end{figure}

The mutually stabilising pairs a particle participates in, and thus the bridge properties, are significantly affected by the choice of the stabilising subset. This is illustrated by a (two-dimensional) example (Fig.~\ref{choiceofsubset}). Applying the LCOM criterion (Fig.~\ref{choiceofsubset}A) results in two one-particle bridges; particle 2 is stabilised by particles 1 and 5, particle 3 is stabilised by particles 4 and 5, with three (independently stabilised) base particles 1, 4 and 5. For the same particle configuration, the LSQS criterion (Fig.~\ref{choiceofsubset}B) considers particle 2 as stabilised by particles 1 and 3, and particle 3 as stabilised by particles 2 and 4, with particles 1 and 4 being base particles and particle 5 not participating at all in this bridge. Particles 2 and 3 are thus a mutually stabilising pair and hence form a two-particle bridge according to the LSQS criterion, but not the LCOM criterion. As an aside, this example also illustrates how a bridge is terminated by non-mutually stabilising particles (which is not to say that they are not otherwise stabilised): Particles participating in only one mutually stabilising pair terminate a bridge, those participating in two mutually stabilising pairs continue bridges, and particles which occur in more than two mutually stabilising pairs lead to branching points and thus a `bridge network'.  Note that Fig.~\ref{choiceofsubset} also illustrates the problem the LSQS criterion addresses, namely that particle 3, being smaller than the mean particle size, is not a genuinely contacting neighbour of particle 3; the requirement that $c>1$ (Step 3) means that it is for the purpose of the bridging analysis considered so.

\begin{figure}[h]
\begin{center}
\includegraphics[width=10cm,angle=0,clip=true,trim=0 0 0 0]{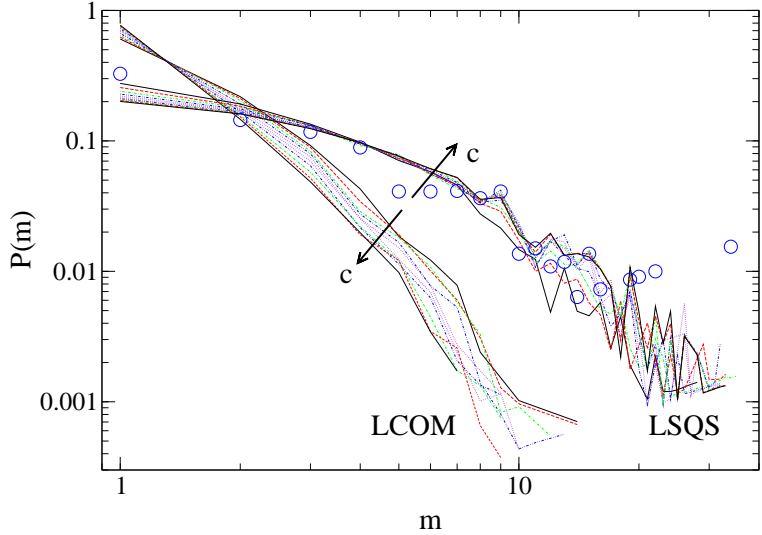}
\caption{Bridge size distribution $P(m)$ for different cutoff values
$c$ in the range $1.05\le c\le1.14$, for the LCOM and LSQS
stabilising subset selection criteria (as indicated).  The colours
and line styles are for clarity only; the evolution with increasing
cutoff $c$ is monotonic in $c$ in both cases.  Sample as in Figure
\ref{coordnowithcutoff}. The blue circles indicate $P(m)$ for a
simulated granular sample (see text).
\label{BridgeParamsbridgedistsboth}}
\end{center}
\end{figure}

We have studied the effect of the choice of the stabilising subset, the LCOM or LSQS criterion, on the bridge size distribution $P(m)$ (Fig.~\ref{BridgeParamsbridgedistsboth}). The LSQS criterion leads to significantly larger bridges than the LCOM criterion.  We attribute this to the fact that if two particles are close to one another, it is likely that they are chosen as mutually stabilising since the short distance between their centres means a small contribution to the mean squared separation {\em for both particles}.  The LSQS criterion, which favours nearby particles, is likely therefore to favour subsets which identify those two particles as mutually stabilising.  This leads to many mutual stabilisations, and in turn many bridges of various sizes.  This is different if the LCOM criterion is used: if a first particle is chosen as stabilising a second, the second is likely to lie higher than that first by the definition of the LCOM criterion; the second is therefore unlikely to be chosen by the same criterion as stabilising the first.

The behaviour of the colloidal samples is compared to a simulation result for a granular system (Fig.~\ref{BridgeParamsbridgedistsboth}). The simulation considered a collection of $2200$ unit-diameter spheres in a (periodic) box of square base size $6\times6$. The spheres were allowed to form a deposit of volume fraction $\phi \approx 0.57$ using a so-called `drop-and-roll' procedure, which captures mutual stabilisations (for details see \cite{Pugnaloni01}). The bridge size distribution $P(m)$ of the simulated granular packing is very similar to the bridge size distribution $P(m)$ determined using the LSQS criterion for all but the highest values of $m$, which reflects the limited system size. This suggests an underlying general bridge size distribution and thus a common structure formed by both the colloidal and granular system.

\begin{figure}[h]
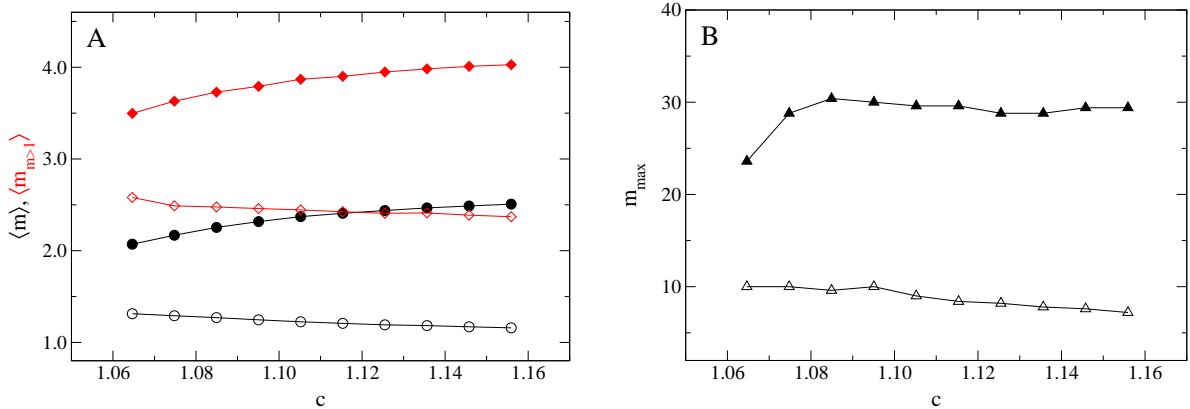

\begin{center}
\mbox{
\includegraphics[width=7.5cm,angle=0,clip=true,trim=0 0 0 0]{meanbridgesizewithcutoff.eps}\qquad
\includegraphics[width=7.5cm,angle=0,clip=true,trim=0 0 0 0]{maxbridgesizewithcutoff.eps}
}
\caption{(A) Mean bridge size $\langle m \rangle$ as a function of cutoff $c$ for all bridges (black circles) and for only bridges of size $m > 1$ (red diamonds).  Results are shown for the LCOM (open symbols) and LSQS (filled symbols) criterion.  (B) Maximum bridge size $m_\mathrm{max}$ for LCOM (open triangles) and LSQS (filled triangles).  Sample as in figure \ref{coordnowithcutoff}.  \label{meanbridgesizewithcutoff}}
\end{center}
\end{figure}

The characteristic properties of the bridge size distribution $P(m)$, namely the mean bridge size $\langle m \rangle$ and maximum bridge size $m_{\mathrm{max}}$, also reflect the dependence on the choice of the stabilising subset (Fig.~\ref{meanbridgesizewithcutoff}).  Consistent with the $P(m)$ results, LSQS leads to significantly larger bridges, in terms of mean as well as maximum size, than LCOM.

\subsection{Choice of the cutoff value $c$}
\label{sec:discCutoff}

It is not {\em a priori} clear which value of cutoff $c$ is most appropriate, beyond that it should lie within a range given by the particle polydispersity and the uncertainty in the co-ordinates (Section \ref{algorithm}, Step 3).  We arrive at a reasonable value by considering how varying $c$ alters the bridge size distribution $P(m)$ and the proportion of stable particles in the packing.

The bridge size distribution $P(m)$ depends on both the cutoff value $c$ and the selection of the stabilising subset. With increasing $c$, $P(m)$ based on the LCOM criterion shows a steeper decrease and thus tends towards smaller bridges (Fig.~\ref{BridgeParamsbridgedistsboth}). With increasing cutoff value $c$, the chosen stabilising particles lie increasingly lower below the stabilised particle. This makes it increasingly unlikely that the upper particle is considered stabilising the lower particle, which tends to break mutual stabilisations and leads to smaller bridges.

In contrast, in the LSQS case a larger cutoff value $c$ leads to only modestly larger bridges; $P(m)$ increases only slightly in a narrow range of $c$ and then reaches some limiting distribution which we call $P_0(m)$.  This is consistent with the fact that, as $c$ is increased further, all additional, more distant neighbours are no longer chosen by the LSQS criterion, which selects the closest stabilising subset, and thus $P(m)$ remains unchanged. Nevertheless, increasing $c$ does include further candidate particles, which, together with the above-mentioned bias toward mutual stabilisations, occasionally leads to bridges with an increased number of particles $m$. This requires that a relatively distant particle can be combined in a stabilising subset with two very close particles, which could not participate in a stabilising subset without the distant particle, resulting in a lower {\em mean} separation squared. Although this is possible, it is very unlikely and hence leads to only a slight increase in bridge size with increasing $c$.

The dependencies on $c$ are also reflected in the mean bridge size
$\langle m \rangle$ and maximum bridge size $m_{\mathrm{max}}$
(Fig.~\ref{meanbridgesizewithcutoff}).  Using the LCOM criterion
results in a slight dependence of $\langle m \rangle$ and
$m_{\mathrm{max}}$ on $c$, while the LSQS criterion leads to
increasing and then, for $c \gtrsim 1.12$, saturating $\langle m
\rangle$ and $m_{\mathrm{max}}$. The weak dependence of the bridge
size distribution $P(m)$ based on the LSQS criterion on the cutoff
value $c$, especially for large $c$, indicates that this criterion
compensates for experimental uncertainties in particle location and
size polydispersity.  This is supported by the agreement between our
experimentally found bridges and the simulated granular bridges
\ref{BridgeParamsbridgedistsboth}.  We thus favour the LSQS
criterion over the LCOM criterion for experimental data, where
variability in the particle size and uncertainty in their locations
are important.  Furthermore, a favourable choice of the cutoff value
$c$ seems to be $c=1.12$.  For this value the number of rattlers,
$1-p_s$ (Fig.~\ref{stablewithcutoff}), is close to its saturation
value and is consistent with other studies \cite{OHern03}.  In
addition, $m_{\mathrm{max}}$ and $\langle m \rangle$ are nearly
constant around this value (Fig.~\ref{meanbridgesizewithcutoff}).


\section{Conclusions}

We have investigated bridging in sedimented packings of colloidal spheres and compared them to simulations of granular packings. While thermal motion dominates in colloidal samples, granular systems are subject to the effect of gravity. Nevertheless, identical bridging behaviour, as judged from the bridge size distribution $P(m)$, was observed in packings formed under both conditions. Despite this agreement in $P(m)$, care should be taken when associating bridges, as found using the algorithm described, with genuine load-bearing structures.

The bridging analysis was described and a new criterion, the lowest
mean squared separation  (LSQS), for the selection of the
stabilising subset proposed.  This choice takes into account the
uncertainties in experimental data. This selection is one of the two
crucial steps in the bridge finding process. First, when assigning
contacting neighbours a capture criterion or cutoff value $c$ has to
be chosen. This accounts for experimental uncertainties in particle
location and particle size. Due to these uncertainties, which
neighbours are considered to be in contact has to be defined rather
than determined. Second, based on the determined contacting
neighbours, the `most stabilising' subset has to be chosen,
typically from a number of potentially stabilising subsets. This
selection would ideally be guided by load-bearing properties and
thus the observation of force networks, which has been achieved for
(deformable) emulsion droplets \cite{BrujicThesis, Brujic03,
Brujic03b}, but not for hard sphere colloids.   In the latter case,
therefore, only geometrical information is available, namely the
particle location and size (which, in addition, are both subject to
experimental uncertainty). In real load-bearing packings,
furthermore, overstabilisation is usual and thus a single
stabilising subset, consisting of the minimum number of particles
required for stability, does not necessarily exist. Despite these
issues, that there is a single stabilising subset is an integral
assumption of this particular bridging analysis, and as such one
subset has to be chosen.  We have addressed these difficulties in
experimental systems, by considering the stabilising subset with the
on-average nearest neighbours and proposed the lowest mean squared
separation (LSQS) criterion. For reasons discussed above, bridges,
as found using the algorithm described and independent of the
applied criterion, do not necessarily represent genuine load-bearing
structures.  This requires detailed further investigation into
bridging in packings which do and packings which do not actually
bear a load; we do this in forthcoming publications.


\section*{Acknowledgements}
We thank Andrew Schofield for supplying the colloidal particles.
This work was funded by the UK Engineering and Physical Sciences
Research Council (EPSRC) and Rhodia Research and Technology
(Rhodia-Centre de Recherches et Technologies d'Aubervilliers).

\bibliographystyle{unsrt}
\bibliography{BibliographicDatabase}

\end{document}